# An Adaptive, Context-Aware, and Stacked Attention Network-Based Recommendation System to Capture Users' Temporal Preference

Jung-Hsien Chiang, *Senior Member, IEEE*, Chung-Yao Ma, Chi-Shiang Wang and Pei-Yi Hao

***Abstract*—Recommendation systems have become more important since the widespread use of the Internet. The large amount of information means that it is difficult for users to discover they really need. However, users' preferences can change over time because of the age and the impact of social networks. Recommendation systems must capture users' preferences and recommend suitable items, which is a difficult challenge. This study proposes a novel recommendation system that adapts to the changing preferences of users. By learning and adapting to users' changing preferences better recommendations are provided.**

**This study uses context factors as additional information to model users' preferences more precisely. For the proposed recommendation system, the context is based on users' interaction items within each defined time interval. This study proposes two novel attention mechanisms for the recommendation system. The contextual item attention module captures contextual information, changing pattern in users' preference and the importance of items. The multi-head attention module extends the diversity of users' preferences and adapts to changing preferences. The recommendation performance is improved using additional item's temporal information to model the contextual item's representation. Experiments compare the proposed algorithm with several state-of-the-art recommendation methods using three real-world datasets. The experimental results demonstrate that the proposed context-aware recommendation model outperforms traditional methods and demonstrate the effectiveness with which contextual information is captured by the attention mechanism.**

## I. Introduction

Recommendation systems are an important element of many modern Internet services, such as E-commerce and social media platforms. Recommendation systems allow individuals to find what they really want when there is an overwhelming number of choices. Therefore, recommendation systems increase profit for platforms and make life more convenient and enjoyable. It is common to model users' preference by using previous user-item interactions in modern recommendation systems. Most recommendation systems assumed that users' preferences are static. However, users' tastes can change over time according to their age or society. It is difficult for recommendation systems to capture a specific temporal pattern or users' changing preferences over time and make an accurate recommendation (Livne, 2019; Raza, 2019).

To model users' preferences over time, many studies use temporal information with session-based recommendation systems (Chung, 2014; Jannach, 2017). Most studies only consider the order of users' interaction items so there is lack of global preferences. Most use Recurrent Neural Networks (RNN) to account for users' sequential behaviors for a sequential recommendation (Hidasi, 2016) but the emphasis is on users' recent behaviors because of the limitations of a RNN. This recommendation strategy might ignore users' long-term concerns or static interests so studies use context-aware recommendation systems (CARS) (Livne, 2019; Raza, 2019). CARS use additional information about contextual features to predict users' preferences. CARS recommend more precisely using additional temporal, location, or mood information.

In terms of contextual information, time is a significant feature because users' preferences can change over time. CARS combine temporal information with additional contextual information to predict users' preferences more accurately. Although CARS use additional contextual information, most previous studies use a RNN to capture users' temporal preferences (Smirnova, 2017) so there is no consideration of users' long-term concerns or static interests.

To address these limitations, this study develops a novel context-aware recommendation system with two attention modules that consider information about items in context and users' preferences. This proposed recommendation system captures temporal patterns and adapts to users' changing preferences rapidly. Using additional contextual information, the proposed approach better assesses users and makes recommendations more precisely because tastes that change over time are identified using an attention mechanism. This study also proposes a novel multi-head attention module that identifies different aspects of users' preferences, such as types of movies or commodities, to represent users better in the recommendation model.

Figure 1 shows a scenario that illustrates how the proposed recommendation system adapts to the user's changing preferences over time and recommends more accurately. The item context attention module calculates the importance of every interaction item within a defined time interval and captures any changes in the pattern of the user's preference. In this case, the user's preference changes over time according with the individual's age. Figure 1 shows that the girl used to purchase commodities that were cute and pastel colored. In adulthood, she bought more mature and textured items and ten months ago, the attention module identified a change in preference and the module began to adapt to these new preferences. The proposed recommendation system has the following main contribution.

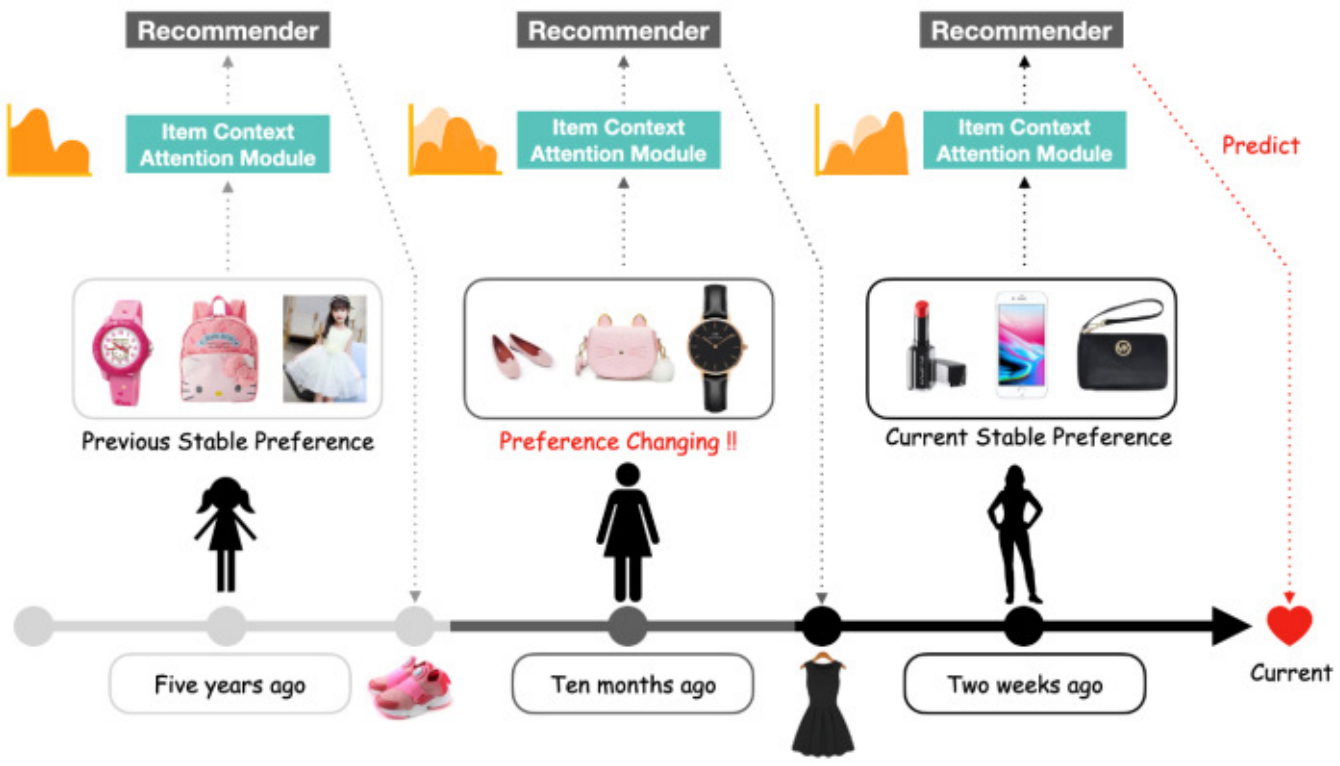

Figure 1. User changes her preference over time and the proposed recommendation system learns to adapt to the changing preference to give a more precise recommendation

- This study proposes a novel item context attention module to capture users' changing preference within a defined time interval. When users interact with items that are different to those they have seen before, the module adapts to the new preference.
- This study proposes a novel multi-head attention module that extends the width of the recommendation model so the proposed model identifies different aspects of users' preferences, such as different types of movies or commodities, and represents users better in the recommendation model.
- In order to capture users' preferences in a higher dimension and to enhance the performance in terms of recommendation, this study proposes a novel stacked structure for a multi-head user attention module to identify fundamental user preferences and to recommend items more precisely.

The remainder of this paper is organized as follows. Section 2 gives a brief review of deep learning-based recommendation systems. Section 3 presents the proposed novel context-aware recommendation system that captures users' changing preferences and identifies different aspects of user preferences. Section 3 also demonstrates how to realize context-aware recommendation that responds to user preferences. Section 4 reports the experimental results. Section 5 draws conclusion and outlines future research.

## II. Background

Deep neural networks have been applied to computer vision, natural language processing and signal processing. This section introduces several approaches that use deep neural networks for a recommendation system.

### *A. Deep Collaborative Filtering based Recommendation Systems*

Deep neural network recommendation systems use classical collaborative filtering (CF) methods. Salakhutdinov et al. (2007) first used Restricted Boltzmann Machines (RBM) for CF to recommend items by modeling the interactions between users and items and calculating the similarities between interactions. He et al. (2017) proposed Neural Collaborative Filtering (NCF), which uses Multilayer Perceptron (MLP) for CF to model the user's interactions with items and calculates the similarities between interactions. Dong et al (2017) used a De-noising Autoencoder (DAE) for CF to extract latent factor vectors for users and items. Guo et al. (2017) proposed a deep learning-based factorization machine (FM) that uses FM for a recommendation system and deep learning for feature learning in a new neural network architecture. Using deep neural networks, CF-based recommendation systems capture more interests and recommend items more precisely. However, these studies do not consider the order and temporal information for items with which users interact. The next section describes several deep neural network approaches that use temporal recommendation systems.

### *B. Deep Session-based Recommendation Systems*

Previous deep learning based CF recommendation systems focus on users' static and global preferences and ignore serial transactional preferences over time so deep session-based recommendation systems were proposed to capture temporal information from items with which users interact. Li et al. (2012) proposed a topic-based hidden Markov model to analyze the temporal dynamics of user preferences.

In recent years, Recurrent Neural Networks (RNN) have been used to capture serial information along a temporal sequence. Zhang et al. (2014) propose a framework that directly models the dependency on users' sequential behaviors into a click prediction using a RNN. Hidasi et al. (2016) proposed a session-based recommendation system that uses parallel RNN architectures to model clicks, images and text features of items with which users interact. Hidasi et al. (2016) proposed a session-based recommendation model, called GRU4REC, which captures users' temporal preferences from user-item interactions. The study used a Gate Recurrent Unit (GRU), rather than a traditional RNN model, to address gradient vanishing. GRU4REC can also be used for data augmentation (Tan, 2016) and with other approaches (Jannach, 2017).

### *C. Deep Attention-based Recommendation Systems*

This section describes recommendation systems with an attention mechanism. Attention mechanisms were first applied in neural machine translation. (Bahdanau, 2014), learned to pay attention to the most important parts of the target. A neural machine translation model with attention mechanism learns which target words contribute more importance to a new translated word and makes long-range translation more fluently. In addition to neural machine translation, attention mechanisms are also applied in image capture (Vinyals, 2019) and speech recognition (Chorowski, 2014). Google Brain uses a self-attention mechanism and the Transformer model architecture (Vaswani, 2017) to give a superior attention mechanism.

Several recent studies use attention mechanisms for a recommendation approach. Gong et al. (2013) proposed a Convolutional Neural Network (CNN)-based recommendation system that uses an attention mechanism to give hash-tag recommendations. Ying et al. (2016) proposed a two-layer hierarchical attention network to capture users' long-term and short-term preferences. He et al. (2018) proposed a deep learning-based item-to-item CF model with an attention mechanism to calculate item similarity for recommendations. Qiu et al. (2018) used a new neural networks architecture, called Graph Attention Networks (Veličković et al., 2017), for social recommendation. Zhang et al. (2018) used an attention mechanism for a sequence-aware approach to learn item-item relationships for recommendations.

An attention mechanism captures the importance of items and then identifies the users' long-term and short-term preferences. This study proposes a novel attention-based, context-aware recommendation system. Using an attention mechanism, the proposed approach combines the strengths of CF approaches and session-based approaches and recommends more precisely using users' long-term and short-term preferences.

### D. Context-Aware Recommendation Systems (CARS)

Most CF-based recommendation systems only consider items with users have interacted, so these recommendation systems focus on users' long-term and static preferences. Session-based recommendation systems consider items with which users have recently interacted and the order of those items, so session-based recommendation systems focus on users' short-term preferences. Many studies model the context and the time of day, the location, the users' devices or the current mood of users. Using contextual factors, recommendation system assess users better and identify any effect on users' preferences. Adomavicius et al. (2005) proposed a multidimensional approach that provides recommendations based on additional contextual information. Tu et al (2017) proposed a Network embedding (NE) model to learn context embedding. Wang et al. (2018) designed an attention-based model for context embedding to weight each interaction item in a transaction without assuming order. DaCosta et al. (2019) combined a deep learning based CF method with collective embedding for CARS. Wang et al. (2019) proposed a time-aware model for user-item ratings that uses local coherence.

Using various contextual factors, recommendation systems identify users' changing preferences in contextual environments. If users' dynamic preferences are better identified, recommendation system assess users better than those that only consider static preferences or short-term preferences. Therefore, the proposed system recommends items more precisely to users using additional information.

The disadvantage of session-based recommendation systems is that the results of a recommendation are significantly affected by the nearest items so they cannot recommend precisely because they cannot identify users' global preferences. Therefore, this study uses information about users and items to identify users' long-term and short-term preferences and to capture changing pattern of preference using user-item attention.

Previous time-aware CARS use a RNN for the recommendation model (Smirnova et.al. (2017) and Wang et al. (2019)). These studies use additional context information to better assess users but they do not address the drawbacks of previous session-based recommendation systems. DaCosta et.al. (2019) used the interaction time point as a collaborative feature in a CF-based model, but this approach does not consider the importance of each interaction item and is affected by outliers and non-important items. Therefore, this study uses an attention mechanism.

## III. Methodology

This study proposes a novel context-aware model that uses two different attention modules to capture user preferences and contextual information about items. For all contextual factors, temporal information is the most important factor in any change in users' dynamic preferences. Therefore, the contextual information for the proposed model uses items with which users have interacted within a defined time interval. The attention mechanism performed well in previous studies in terms of capturing dynamic changes in the patterns of users' preferences without specifying an order. Therefore, this study uses two attention modules that enable the proposed context-aware recommendation model to rapidly adapt to changes in users' preferences.

Figure 2 shows the proposed context-aware model, called a *Contextual User Preference Attention Model*. This model comprises two attention modules: the *Multi-Head User Attention Module* and the *Contextual Item Attention Module*. The user and item embedding is firstly input into the contextual item attention module to represent the contextual item, which is the concentrated item representation of the current context. The context item attention module combines the advantages of an attention mechanism and a context-aware recommendation system to adapt to users' changing preferences and recommend more precisely. The contextual item attention module captures users' preference patterns from the items with which users interact within each defined time interval using an attention network. A multi-head user attention module then identifies different aspects of users' preferences and restructures new precise users' preferences using an attention network. The restructured users' preferences are used to calculate the scores for each candidate item using a preference scoring function. These scores are used to rank the candidate items and to select the most suitable recommended items. The notation that is used in this manuscript is summarized in Table I.

TABLE I

| Notation | Description |
|---|---|
| $U,I$ | user and item set |
| $C_u$ | set of items with which a specific user *u* has recently interacted in a defined time interval |
| $k \in N$ | dimensionality of user and item embedding vector |
| $t \in N$ | number of items in a given time interval |
| $m \in N$ | number of Multi-head User Attention Modules in stacked structure |
| $h \in N$ | number of heads in multi-head user attention module |
| $p_j \in R^k$ | item embedding vector for the *j*th item |
| $q_i \in R^k$ | user embedding vector for the *i*th user |
| $\alpha_t \in R$ | attention weights of items' representation vectors $p_t$ in *Contextual Item Attention Module* |
| $s_t \in R^k$ | weighted items' representation in *Contextual Item Attention Module* |
| $C \in R^k$ | temporal contextual item representation in *Contextual Item Attention Module* |
| $PE_{pos} \in R^k$ | positional encoding vector in *Contextual Item Attention Module* |
| $y^m \in R^k$ | user preferences representation in the *m*-th *Multi-head User Attention Module* |
| $w_n^m \in R^{k \times k}$ | matrix of projected spaces to identify different aspects of a user in the *m*-th *Multi-head User Attention Module* |
| $d_n^m \in R^k$ | projected vector of user aspects in the *m*-th *Multi-head User Attention Module* |
| $\beta_n^m \in R$ | attention weights in the *m*-th *Multi-head User Attention Module* |

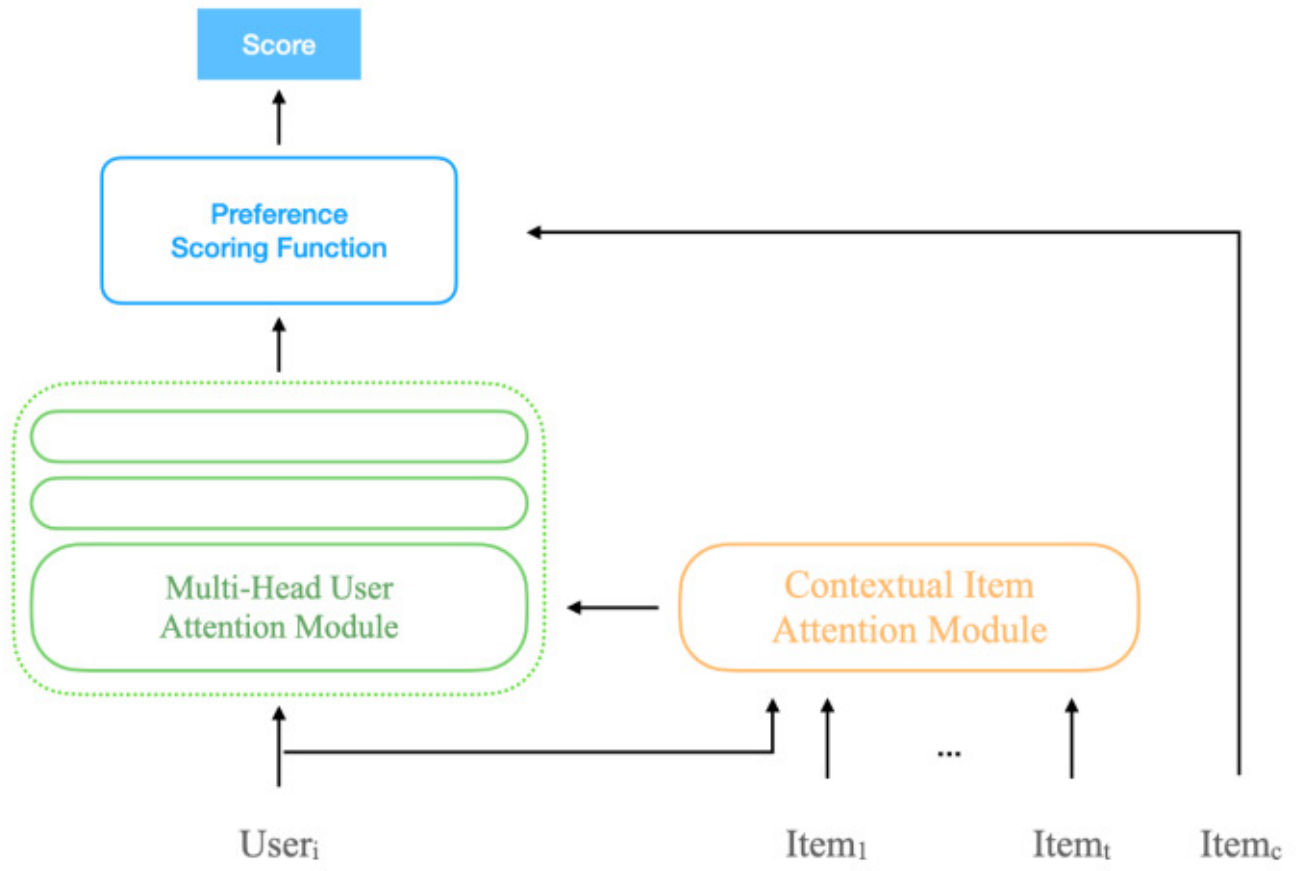

Figure 2. Overview of Contextual User Preference Attention Model

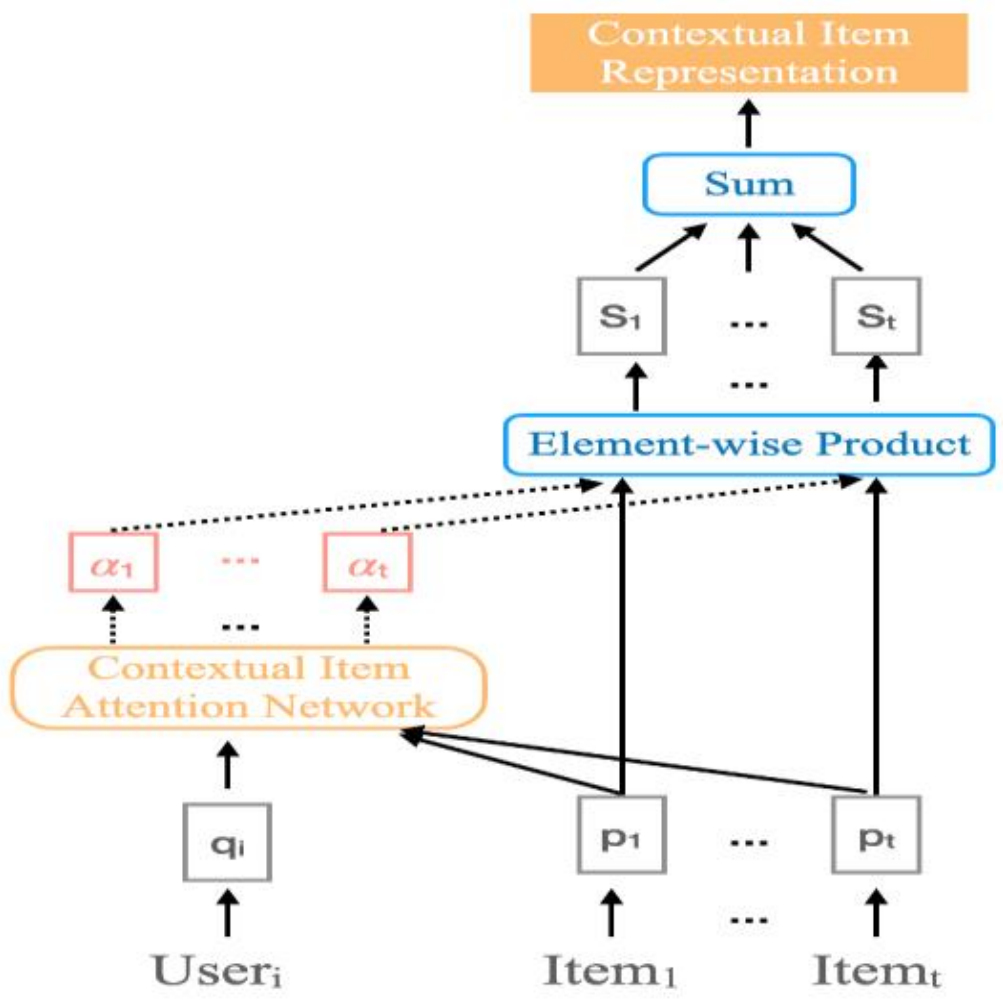

Figure 3 Contextual item attention module

### A. Contextual Item Attention Module

The contextual item attention module models contextual information by embedding users and embedding items with which a specific user interacts within a defined time interval. Temporal information is primarily related to users' changing preferences and is more easily derived from existing data so this study uses a time factor as a feature to contribute the contextual information for the proposed context-aware recommendation system.

#### 1) Item Context Modeling using a Contextual Item Attention Module

For deep learning based recommendation systems, user embedding and item embedding are the most common techniques that are used to represents users and items using $k$-dimensional vectors. If $U$={User$_1$, User$_2$, …, User$_{|U|}$} denotes a set of all users, the $k$-dimensional user representation vector $q_i \in R^k$ for $1 \le i \le |U|$ denotes the user embedding for any User$_i$ in set $U$. Let $I$={Item$_1$,Item$_2$,…,Item$_{|I|}$} denotes a set of all items, the $k$-dimensional item's representation vector $p_j \in R^k$ for $1 \le j \le |I|$ denotes the item embedding for any Item$_j$ in set $I$. For the proposed context-aware recommendation, a user's interaction sequence $C_u = (Item_1, Item_2, \ldots, Item_t)$, is used to predict the next item. During the training process, the model predicts the next item depending on the previous $t$ items and the user' preference. Figure 3 shows that the inputs for the proposed contextual item attention module are the user's representation vectors $q_i$ and item representation vectors $p_1, \ldots, p_t$ within a defined time interval. The user's interaction sequence $C_u = (Item_1, Item_2, \ldots, Item_t)$ represents the $t$ items with which a specific user has recently interacted within a defined time interval, which provides the contextual information for the user. Information about a user 'schanging preference over time is also incorporated. The proposed item context attention module captures contextual information and users' changing preferences within a defined time interval. The attention weights $\alpha_t \in R$ are calculated as:

$$\alpha_t = \mathrm{softmax}(q_i^T p_t) \quad (1)$$

where $\alpha_t$ denotes the attention weights of item representation vectors $p_1, \ldots, p_t$ within the defined time interval. The weighted items' representation $s_t$ are calculated using the formula:

$$s_t = \alpha_t p_t \quad (2)$$

Finally, the temporal contextual item representation between user and items $C$ that is modeled using the attention mechanism is calculated as:

$$C = \sum_t s_t \quad (3)$$

#### 2) Item Embedding with Positional Encoding

This section incorporates positional encoding into the proposed Contextual Item Attention Module to improve the performance of the proposed system by using temporal information about additional items. Positional encoding was first used for natural language processing (NLP) research (Vaswani, 2017) by Google Brain and gives extra positional information for words in a sentence. This study uses positional encoding as time information for each item. The positional encoding vector $PE_{pos}$ is calculated as:

$$PE_{(pos,2i)} = \sin(pos/10000^{2i/k}) \quad (4)$$
$$PE_{(pos,2i+1)} = \cos(pos/10000^{2i/k}),$$

where $pos$ denotes the position, which is the order of items, and $i$ is the index of dimension. $2i$, which is the index of a dimension for which the index is even and $2i$+1 denotes the index of a dimension where the index is odd, and $k$ is the total length as the embedding vector. Using this positional encoding algorithm, vectors with the same dimension and amount of input items are generated. Positional encodings are then added to the input item embedding $p_t$ to give additional positional information for the attention model. As shown in Figure 4, the positional encoding is calculated using the above formulation and the positional encoding vector is added to the items' embeddins $p_t$, so:

$$p_t = p_t + PE_t \quad (5)$$

The positional encoding has the same dimension $k$ as the items embedding, so the two vectors can be summed. This function is chosen because the model can easily learn to attend using relative positions, because for any fixed offset $k$, $PE_{pos+k}$ is represented as a linear function of $PE_{pos}$ (Vaswani, 2017).

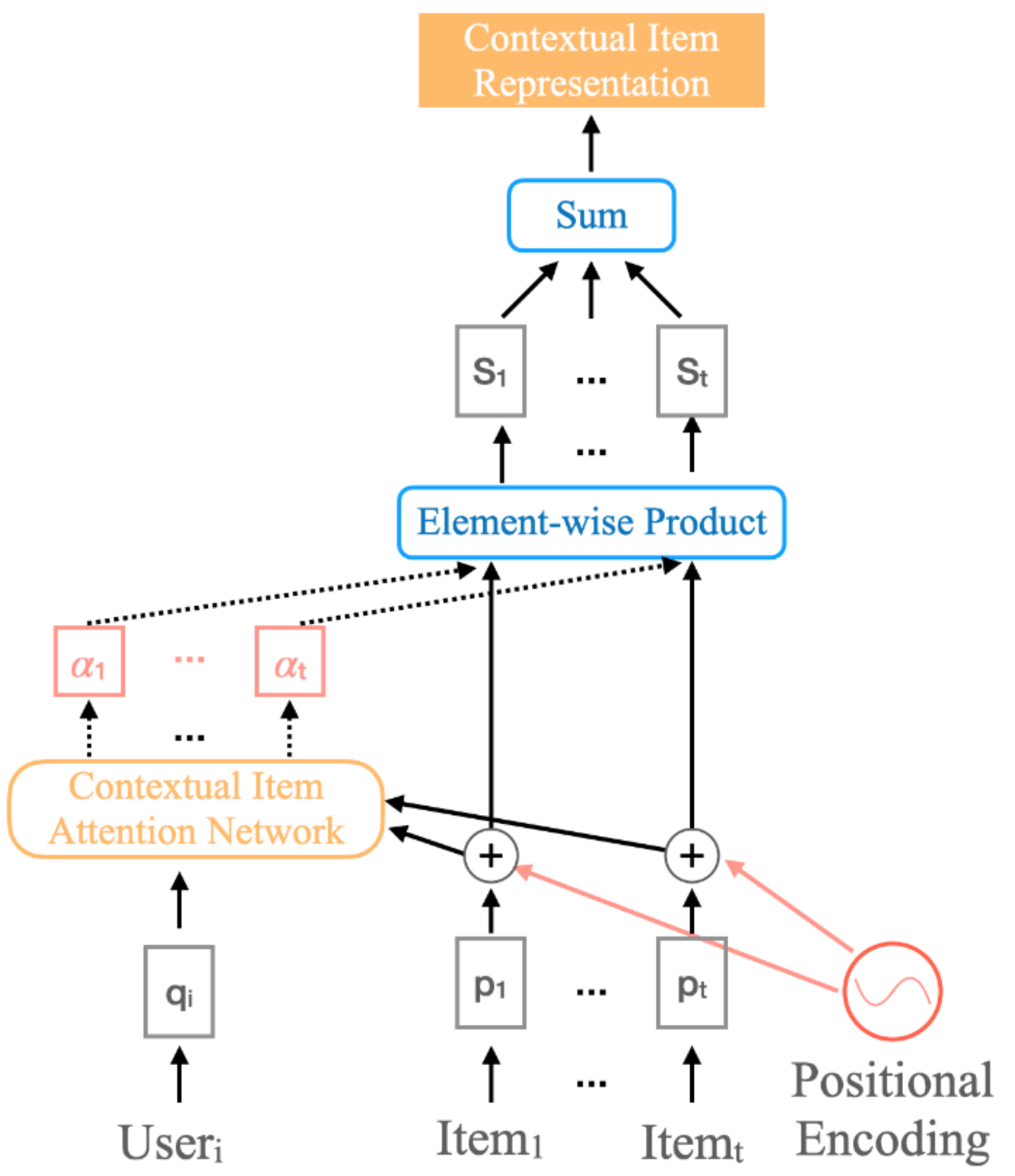

Figure 4 Contextual item attention modules with positional encoding

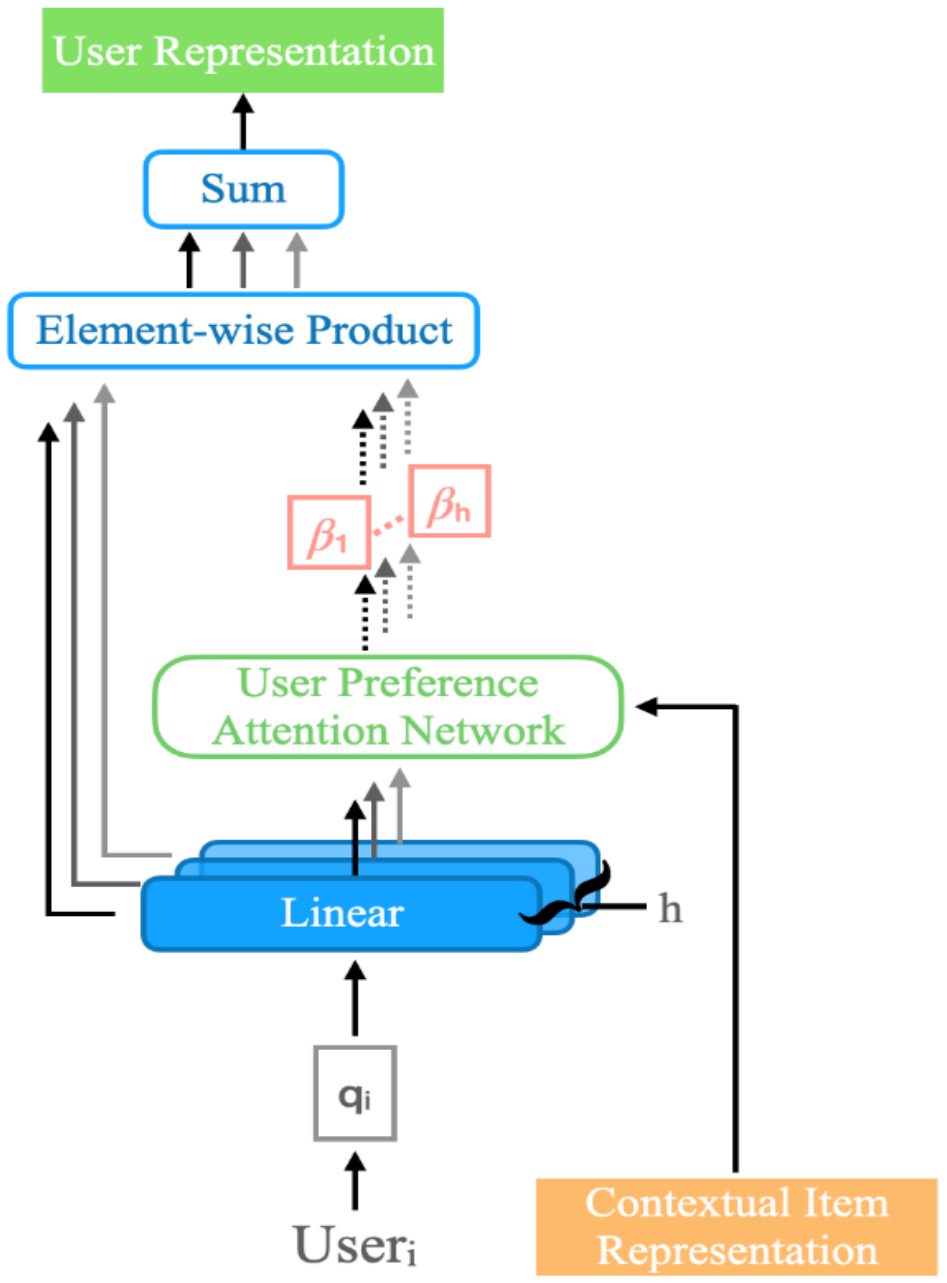

Figure 5 Multi-Head User Attention Module

## B. *Multi-head User Attention Module*

Inspired by the approach of Vaswani et al. (2017), the proposed model uses a multi-head attention mechanism to identify different aspects of users' tastes and to extend the diversity of users. In order to capture users' preferences in a higher dimension and to improve the recommendation, this study proposes a novel stacked structure with a multi-head user attention module to identify fundamental user preferences.

### 1) *User Preference Modeling with a Multi-head User Attention Module*

As shown in Figure 5, the input of the Multi-Head User Attention Module is the $k$-dimensional user's representation vector $q_i$ and the contextual item representation $C$, where the contextual item representation $C$ is the output of the contextual item attention module. The user embedding $q_i$ is linearly projected onto different projected spaces $W_n$ in parallel $h$ times to identify different aspects of a user with the same $k$ dimension. The projections $d_n \in R^k$ for $1 \leq n \leq h$ are calculated as:

$$d_n = W_n q_i . \tag{6}$$

An attention mechanism then links the contextual item representation $C$ and projections to model user preferences $y$. The user preferences $y$ are calculated as:

$$y = \sum_{n=1}^{h} \beta_n d_n \tag{7}$$

where the attention weights $\beta_n \in R$ are calculated as:

$$\beta_n = C^T d_n . \tag{8}$$

For user preferences $y$, the output of a multi-head user attention module calculates the scores for an item and rank items according to the score to build a recommendation list for a given user. The preference scoring function is described in Section III-C.

### 2) *Stacked Multi-head User Attention Module*

User preference $y$ essentially aggregates a user's representation vector $q_i$ and the contextual item representation $C$. However, more complex user preferences are identified using another Multi-Head User Attention Module. In order to capture users' preferences in a higher dimension and to enhance the performance of recommendation, this study proposes a novel stacked structure with a multi-head user attention module to identify fundamental user preferences and to recommend precisely. In Figure 6, for the first multi-head user attention module in the first layer of the stacked structure, the user's embedding and contextual item representation $C$ are the input, similarly to the normal multi-head user attention module. However, for the second layer multi-head user attention module, the contextual item representation $C$ and the users' preferences $y$ are used as inputs, where $y$ is the output from the multi-head user attention module in the previous layer in the stacked structure. User embedding is replaced by a user preferences representation, which is the output from the previous module. The output of the $m$-th multi-head user attention module is calculated as:

$$y^m = \sum_{n=1}^{h} \beta_n^m d_n^m \tag{9}$$

where $y^m$, $m \geq 2$ is the user preferences representation output from the *m-th* multi-head user attention module. $\beta_n^m$ is calculated as:

$$\beta_n^m = C^T d_n^m , \tag{10}$$

and $d_n^m$ is calculated as:

$$d_n^m = W_n^m y^{m-1} , \tag{11}$$

where $y^{m-1}$ is output of the ($m$-1)-th multi-head user attention module in the stacked structure. The proposed stacked structure identifies users' fundamental preferences by projecting user representation repeatedly into different linear spaces. User preferences $y^m$, are used to calculate the preference scores for a user-item pair, and the item with the highest score is recommended to a specific user.

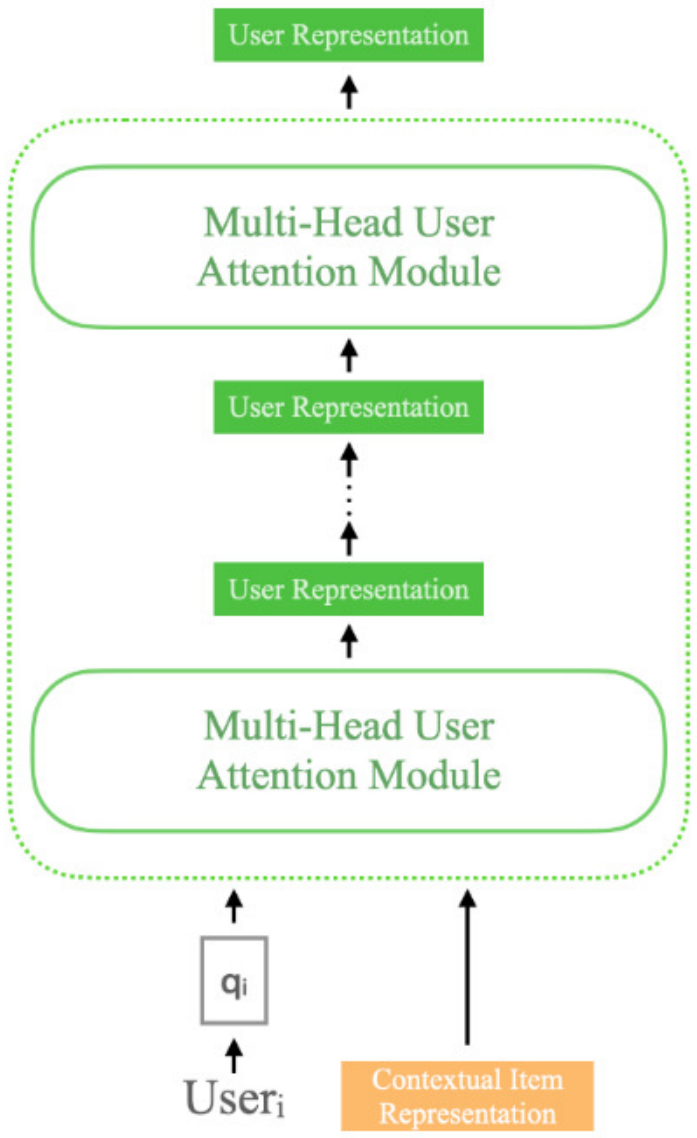

Figure 6. Stacked multi-head user attention module

### C. Recommendation System with a Contextual User Preference Attention Model

This section uses the proposed Contextual User Preference Attention Model for a recommendation system. Training techniques and the learning objective for the proposed recommendation system are also detailed. The score and rank of recommended items for the proposed recommendation system are also described.

#### 1) Pre-training and Item Embedding

Pre-training is a common technique of enhancing performance and accelerating convergence for a deep learning based model. This study uses the pre-trained weights from a Bayesian Personalized Ranking (BPR) (Rendle, 2009) method as the initial values for user and item embedding for the proposed model. BPR uses user-item interactions with their embedding to predict the rating for items. When BPR is complete, the embedding weight of users and items that are obtained from the pre-trained BPR area assigned as the initial values for user and item embedding for the proposed model. This user and item embedding is finely tuned during the learning phase that is described in next Section.

#### 2) Scoring Function and Model Training

This section uses the Contextual User Preference Attention Model for a context-aware recommendation system and recommends items to users using a preference scoring function. Items are initially ranked in terms of their preference scores, which are calculated as:

$$score = \sigma(y_i^T p_c) \qquad (12)$$

where $y_i$ is the user representation of User$_i$ and $p_c$ is the embedding of candidate Item$_c$. The user preference representation $y_i$ is the output of the proposed *Multi-Head User Attention Module*. The item embedding $p_c$ is learned using a pair-wise learning method. When the scores of candidate items are calculated, they are ranked in descending order and the item with the highest score is recommended.

This study uses a pairwise learning objective method during the learning phase of the proposed recommendation system. The pairwise learning objective method uses BPR (Al et al., 2009), which uses implicit feedbacks, such as purchase records or clicks (e.g. click or not click) because users usually provide implicit feedback instead of explicit feedback (e.g. ratings from 1 to 5) for an existing dataset. The pairwise learning objective method assumes that items with which a user has interacted before are rated as 1 and those with which a user has not interacted are assumed to be rated as 0. If *Item$_j$* is selected by *User$_i$*, then the rating $r_{ij}$ between *User$_i$* and *Item$_j$* is rated as 1. The objective of the pairwise learning objective method is to make the rating for the positive items set, $Item_j^+$, that is the interaction items, higher than the rating for the set of negative items, $Item_k^-$, which are the items with which a user has not been interacted. For an embedding pair $(q_i, p_j^+, p_k^-)$ from the defined training triple $(User_i, Item_j^+, Item_k^-)$, the objective function for the pairwise learning method in a traditional BPR algorithm is:

$$\min \sum\nolimits_{(i,j,k)\in D} -\log(\sigma(q_i^T p_j^+ - q_i^T p_k^-)) \qquad (13)$$

User embedding $q_i$ is replaced with user preference representation $y_i$, which is the output of the proposed *Stacked Multi-head User Attention Module*, so the pair wise objective function for the proposed context-aware recommendation model is:

$$\min \sum\nolimits_{(i,j,k)\in D} -\log(\sigma(y_i^T p_j^+ - y_i^T p_k^-)) \qquad (14)$$

This study uses the same strategy as (He, 2017) to reduce the computational cost by randomly sampling 10 negative items from those that have not been bserved by each of the users, instead of all of the unobserved items in training. During the training phase, parameters are learned and updated by the proposed commendation system using the Adam (Kingma, 2015) algorithm.

### D. Complexity Analysis

#### 1) Time Complexity

The computational complexity of the proposed model depends on the *Item Context Attention Module* and *Stacked Multi-head User Attention Module*, which is O($h\times m\times k^2 + t\times k^2$), where $t$ denotes the number of items within a defined time interval, $h$ denotes the number of heads and $m$ denotes the depth of the *Stacked Multi-Head User Attention Module*. A convenient property of the proposed model is that the computation for each attention model is fully parallelizable, so GPU acceleration is possible. In contrast, RNN-based methods (Chung, 2014) use time steps (i.e., computation on time step $t$ must wait for results from time step $t$-1), so there are O($t$) time on sequential operations. The proposed method is more than five times faster than RNN -based methods with GPUs. During testing, when the user embedding and item embedding are calculated, the process is the same as that for standard MF methods (O($k$) for evaluating preference for an item).

*2) Space Complexity*

The learned parameters for the proposed model are from the user embeddings and item embedding and parameters in the *Item Context Attention Module* and the *Stacked Multi-head User Attention Module*. The total number of parameters is $O(|I|\times k + |U|\times k + k^2\times h\times m)$, where $k$ denotes the dimension of user/item embedding vector, $|I|$ and $|U|$ denotes the number of users and items.

## IV. Experiments

This section conducts experiments to answer the following research questions:

**RQ1:** Does the proposed recommendation algorithm outperform state-of-the-art implicit recommendation methods?

**RQ2:** Do deeper layers of stacked structure allow learning from user–item interaction data?

**RQ3:** Is a wider structure for the multi-head attention module better for learning from user–item interaction data?

**RQ4:** What is the effect of the length of the time interval for the proposed architecture?

**RQ5:** What is the effect of the position encoding strategy for the proposed architecture?

### *A. Datasets*

The methods are evaluated using three datasets from three real world applications. These datasets vary significantly in terms of domains, platform, and sparsity:

- **MovieLens:** This is a widely used benchmark dataset for evaluating recommendation algorithms. This study uses the version (MovieLens-100k) that contains 100k user ratings, where each user has at least 20 ratings. This is explicit feedback data so the performance in terms of learning from an implicit signal for explicit feedback is determined. Therefore, the data is transformed into implicit data, where each entry is marked as 0 or 1 to indicate whether the user has rated the item.
- **Pinterest**: This is an image-centered social media dataset that is published by the *Pinterest* platform. Pins (images) are treated as items and boards as users. The original data is very large but highly sparse. More than 20% of users have only one pin, so it is difficult to evaluate recommendation algorithms.
- **Beauty:** This is a series of datasets [McAuley, 2015] comprising large corpora of product reviews from Amazon.com. Top-level product categories on Amazon are treated as separate datasets. The category of Beauty is used for this study. This dataset is notable for its high sparsity and variability.

For *MovieLens*, *Pinterest* and *Beauty* dataset, users whose have less than 10 interactions are filtered. The statistics for each dataset are shown in Table II.

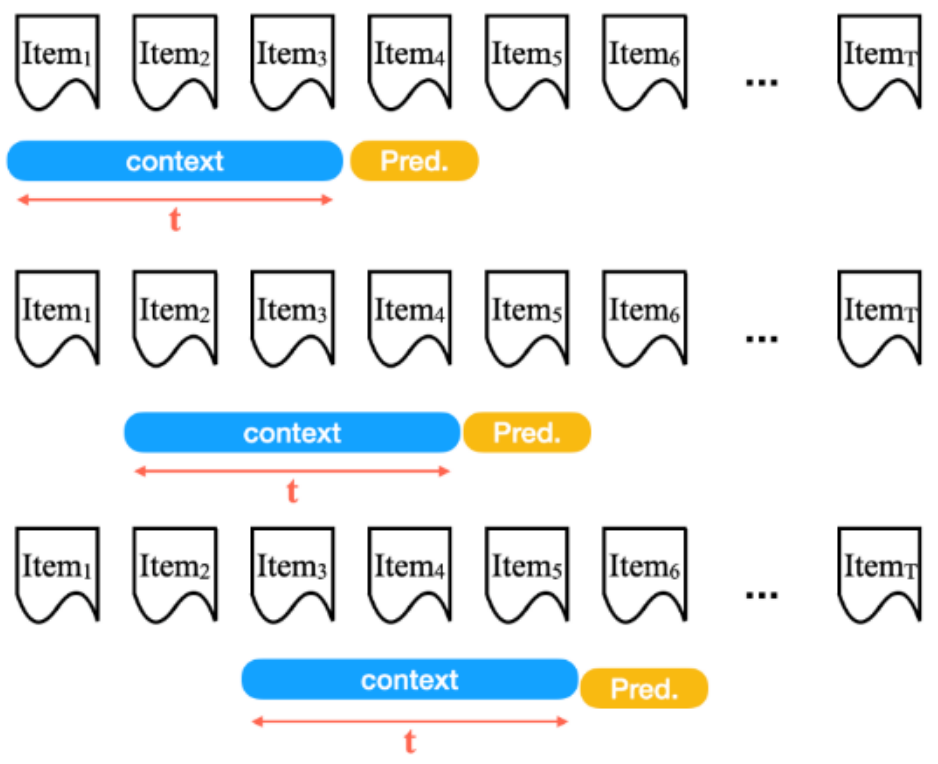

Figure 7. Concept of Input Sequence of Data

TABLE II
STATISTICS OF THREE EVALUATION DATASETS

| Dataset | Interactions | Users | Items | Sparsity |
|---|---|---|---|---|
| MovieLens-100K | 100,000 | 932 | 1,682 | 93.62% |
| Pinterest | 532,203 | 24,935 | 38,756 | 99.94% |
| Beauty | 79,861 | 4,412 | 8,670 | 99.79% |

This study constructs the context according to the items with which a specific user interacts within a defined time interval. The items are separated into many item sequences and the defined time interval is assumed to have a fixed length t. As shown in Figure 7, the input sequence $[Item_j,\ldots,Item_{j+t}]$ from dataset is used to predict $Item_{j+t+1}$. For this experiment, the length of the time interval (window size) for the proposed item context attention module is five.

### *B. Evaluation Metrics*

To evaluate the performance of the proposed recommendation algorithm, a leave-one-out evaluation is used, which is the most common evaluation protocol (Bayer, 2017; He, 2017). For each user, the latest interaction is used as the testing data, which is the ground truth, and the remaining data is used for training. It is costly in terms of time to rank all items for every user during evaluation so a common strategy (Bayer, 2017; He, 2017) randomly samples 100 items with which the user has not interacted, ranking the test items among the 100 items. For a specific user, the recommendation algorithm produces a ranked list of items. To assess the performance of a ranked list with the ground-truth item set (GT), three commonly used evaluation metrics are used: Hit Ratio (HR), Normalized Discounted Cumulative Gain (NDCG) and Mean Reciprocal Rank (MRR). The HR intuitively measures whether the test item is present in the top-k ranked list. If a test item appears in the recommended list, it is deemed a hit. The value for HR@k is calculated as:

$$HR@k = \frac{number\ of\ Hits\ at\ k}{|U|}\ . \qquad (15)$$

The HR is a recall-based metric so it does not reflect the accuracy of top ranks, which is crucial for many real-world applications, so this study also uses Normalized Discounted Cumulative Gain (NDCG), which assigns higher importance to results in the top ranks, scoring successively lower ranks with marginal fractional utility. The NDCG@k is calculated as:

TABLE III
TOP-K RECOMMENDATION PERFORMANCE AT K = 10 FOR THE PROPOSED MODEL AND BASELINE METHODS.

| Dataset | metrics | (a) BPR | (b) NCF | (c) NAIS | (d) GRU | (e) SASRec | Our model | Improvement vs (a)-(e) |
|---|---|---|---|---|---|---|---|---|
| MovieLens-100k | HR@10 | 0.6861 | 0.6532 | 0.6638 | 0.7256 | 0.6957 | **0.7603** | 4.7822% |
| | NDCG@10 | 0.4107 | 0.3803 | 0.3863 | 0.4387 | 0.4166 | **0.4747** | 8.2060% |
| | MRR@10 | 0.3261 | 0.2969 | 0.3015 | 0.3499 | 0.3308 | **0.3861** | 10.3458% |
| Pinterest | HR@10 | 0.8688 | 0.875 | 0.8663 | 0.8415 | 0.9017 | **0.9046** | 0.3216% |
| | NDCG@10 | 0.6199 | 0.6194 | 0.5954 | 0.5633 | 0.6348 | **0.6450** | 1.6068% |
| | MRR@10 | 0.5405 | 0.5384 | 0.5096 | 0.4754 | 0.5493 | **0.5619** | 2.2938% |
| Beauty | HR@10 | 0.5338 | 0.5174 | 0.5274 | 0.4337 | 0.5247 | **0.5587** | 4.6646% |
| | NDCG@10 | 0.2952 | 0.2893 | 0.2985 | 0.2344 | 0.2944 | **0.3132** | 4.9246% |
| | MRR@10 | 0.2222 | 0.2193 | 0.2285 | 0.1736 | 0.2241 | **0.2382** | 4.2450% |

$$NDCG@k = \frac{1}{|U|}\sum_{i \in test\ set}\frac{2^{r_i}-1}{\log_2(i+1)}. \quad (16)$$

Where $r_i$ is the graded relevance of item at position $i$. Simple binary relevance is used so $r_i = 1$ if the item is in the test set and 0 otherwise.

The reciprocal rank of a relevant item response is the multiplicative inverse of the rank of the first correct answer: 1 for first place, 1⁄2 for second place, 1⁄3 for third place and so on. The mean reciprocal rank (MRR) is the average of the reciprocal ranks of results for a relevant item:

$$MRR@k = \frac{1}{|U|}\sum_{i \in test\ set}\frac{1}{rank_i}, \text{where } rank_i < k, \quad (17)$$

where $rank_i$ refers to the rank of item $i$ in the testing set in the top-k ranked list. For all metrics, larger values indicate better performance. The ranked list is limited to 10 for these metrics. The metrics are calculated for each test user and the average score is reported.

## C. Baselines

Using the evaluation metrics, the proposed method is compared to the following baselines.

- **Bayesian Personalized Ranking (BPR) (Rendle, 2009):** BPR optimizes the latent factor of users and items in a recommendation model using a pairwise loss. BPR maximizes the distance between the score for positive items with which the user has interacted r before and the negative items with which the user has not interacted before. Many studies use the training technique for BPR so BPR is a classic baseline for a performance comparison involving recommendation systems.
- **Neural Collaborative Filtering (NCF) (He, 2017):** NCF is a state-of-the-art collaborative filtering-based recommendation system that uses matrix factorization with a deep learning-based method. NCF uses multi-layer perceptron (MLP) instead of the inner product when calculating the relationship between users and items.
- **Neural Attentive Item Similarity Model for Recommendation (NAIS) (He, 2018):** NAIS is one of the state-of-the-art recommendation systems that uses a deep learning strategy for a collaborative filtering recommendation system. Similarly to a Factored Item Similarity Model (FISM) (Kabbur, 2013), NIAS uses two types of item embedding: one to describe the users and another for the items. NAIS improves the FISM model by adding an attention mechanism to model the weight of items that contributes to user representation.
- **Session-based Recommendation System using a Gated Recurrent Unit (GRU) (Chung, 2014):** GRU is a state-of-the-art session-based recommendation system that captures temporal information to recommend items to users. For this experiment, the length of a session for GRU is the same as the length of the time interval (context window size) for the proposed context-aware recommendation system.
- **Self-Attentive Sequential Recommendation (SASRec) (Kang, 2018) :** SASRec is a state-of-the-art recommendation system that is also inspired by (Vaswani, 2017). SASRec uses a self-attention mechanism with an encoder-decoder to replace the Markov Chains and Recurrent Neural Networks of traditional session-based recommendation methods. Using the self-attention encoder-decoder model, SASRec captures both long-term and short-term semantics.

## D. Parameter Settings and Implementation Detail

The hyper parameters for the proposed context-aware recommendation system include the embedding size, the length of the context items sequence, the number of head in multi-head user attention modules and the layers of the stacking structure for multi-head user attention module. The users' embedding and the item embedding are 32-dimensional. For the hyper parameters in stacked multi-head user attention module, the amounts of stack are tested with the values [0, 1, 2, 3] and all with [1,2,3..6] different heads. The proposed recommendation system is implemented using PyTorch and accelerated using a Graphic Processing Unit (Tesla V100). The recommendation model is taught using an Adam optimization algorithm (Kingma &Ba, 2015) with a learning rate of 0.001.

## E. Experimental Results and Analysis

Table III shows the recommendation performance for all methods using the three datasets (**RQ1**). The best results in each row are denoted in bold. As shown in the table, the proposed algorithm achieves the best performance for all datasets, significantly outperforming the state-of-the-art methods by a large margin (The relative improvement over the strongest baseline is 3.25%, 4.91% and 5.63% for HR@10, NDCG@10 and MRR@10, respectively). One likely reason is that the proposed model captures contextual information and changes in user preferences within a defined time interval. When users interact with items that are different to those they have seen before, the proposed module adapts to the change. The stacked multi-head attention module identifies different aspects of users' preferences, and learns to represent users' preferences better for the recommendation model. The proposed model adaptively attends items within the defined time interval within different ranges on different datasets (e.g. only the previous item on sparse datasets and more on dense datasets).

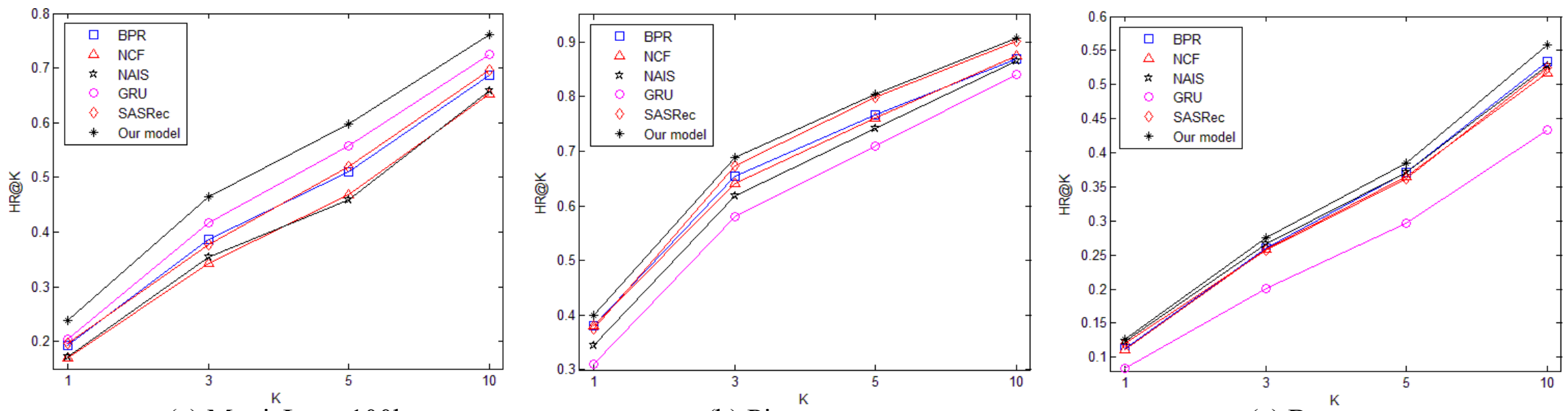

(a) MovieLens-100k (b) Pinterest (c) Beauty

Figure 8. Performance evaluated in terms of the Hit Ration from position 1 to 10 (i.e., *k*)

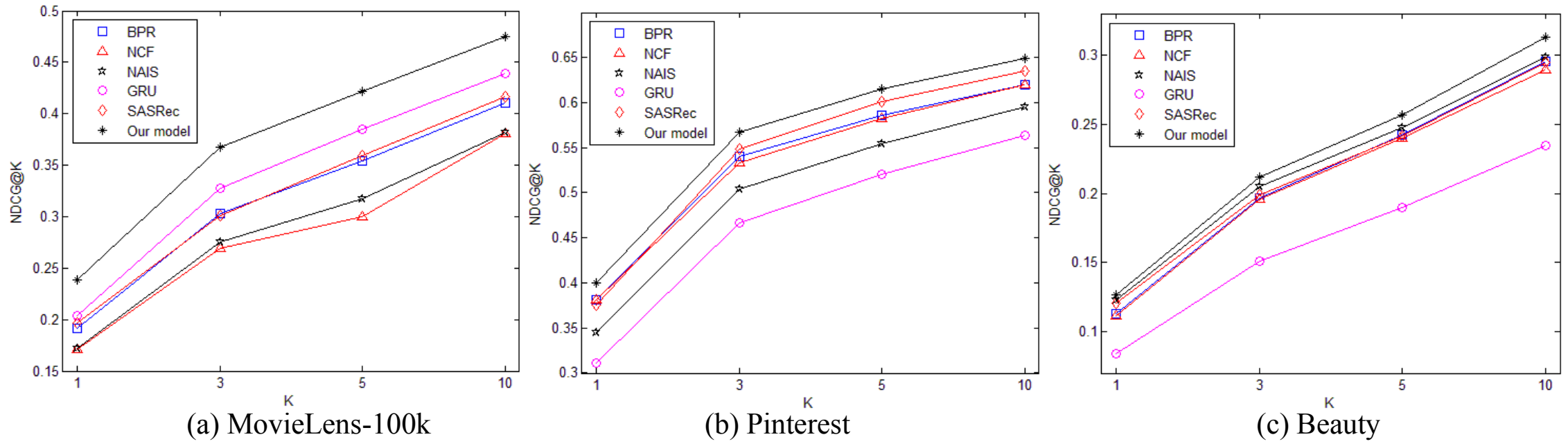

(a) MovieLens-100k (b) Pinterest (c) Beauty

Figure 9. Performance evaluated in terms of NDCG from position 1 to 10 (i.e., *k*)

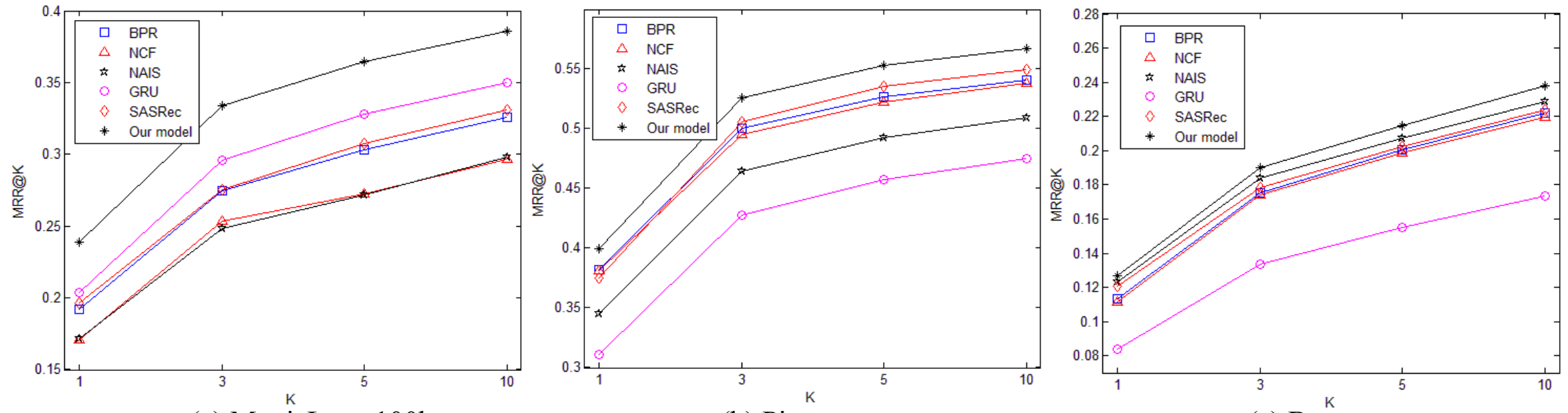

(a) MovieLens-100k (b) Pinterest (c) Beauty

Figure 10. Performance evaluated in terms of MRR from position 1 to 10 (i.e., *k*)

In terms of the evaluation metrics for HR@10, NDCG@10 and MRR@10, the session-based recommendation algorithm GRU achieves the second best recommendation performance using the *MovieLens-100k* dataset, for which the degree of sparsity is 93.62%. However, the recommendation performance of GRU is poorest for the *Pinterest* and *Beauty* datasets, for which the degree of sparsity is 99.94% and 99.79%, respectively. This demonstrates that GRU performs better with denser datasets than sparse datasets, because the GRU approach uses a recurrent neural network, so long user-item interaction lists are required to determine the relationship between users and items. For this dataset, which is extremely sparse, the user-item interaction lists are usually short, so the GRU cannot train the model well and cannot recommend the item accurately. The proposed approach performs well using both sparse and dense datasets.

Figures 8 to 10 show the performance of the Top-*k* recommended lists where the ranking position *k* ranges from 1 to 10. The proposed algorithm demonstrates consistent improvements over other methods across positions.

Figure 11 shows the recommendation performance of the proposed approach using the three datasets, where the number of layers in the stack structure for a multi-head user attention module ranges from 1 to 5 (**RQ2**). Figure 11 shows that the the number of stack layers has little effect on recommendation performance. For the *MovieLens-100k* dataset, the recommendation performance increases slightly as the number of layers in the stack structure increases, because the movie types and user tastes are more complex so it is necessary to determine the fundamental user preference using a deeper stack structure. For the *Pinterest* dataset, the recommendation performance decreases slightly as the number of layers in a stack structure increases because user preference has little effect and the deeper stack structure means that a large number of weights must be learned, which can easily result in overfitting, especially for datasets that are extremely sparse.

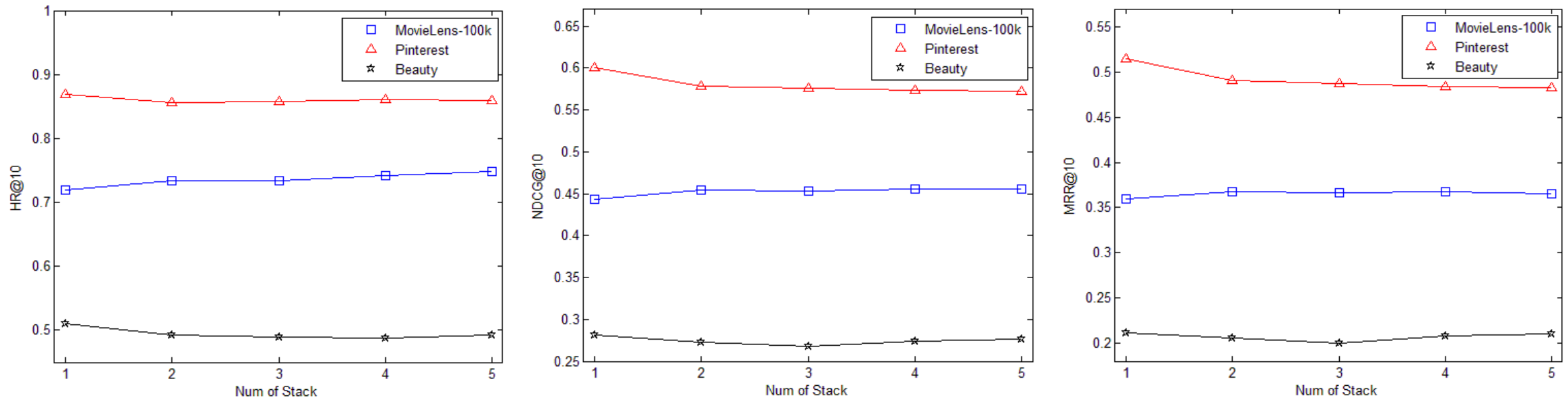

(a) HR@10 (b) NDCG@10 (c) MRR@10

Figure 11. Effect of the depth of stack structure on ranking performance

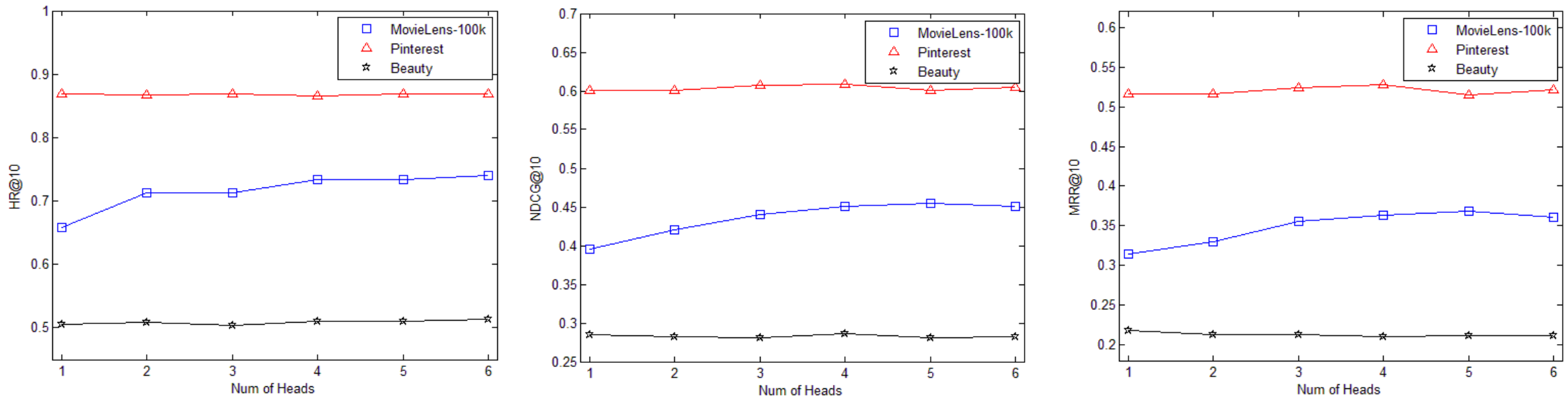

(a) HR@10 (b) NDCG@10 (c) MRR@10

Figure 12. Effect of the number of heads on ranking performance

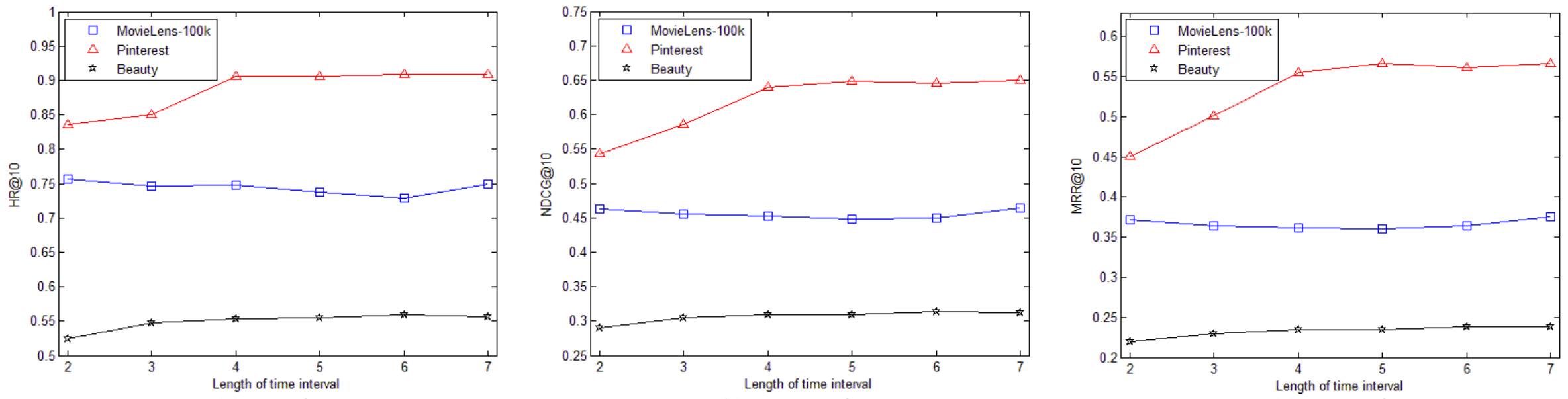

(a) HR@10 (b) NDCG@10 (c) MRR@10

Figure 13. Effect of the length of the time interval (window size) on ranking performance

Figure 12 shows the recommendation performance for the proposed approach using the three datasets, where the number of heads in the multi-head user attention module ranges from 1 to 6 (**RQ3**). As shown in figure 12, for the *MovieLens-100k* dataset, the recommendation performance significantly increases as the number of heads increases because the movie types and user tastes become more complex and more aspects of users' preferences must be identified, such as different types of movies. To represent users better in the recommendation model, more heads are used for the proposed multi-head user attention module. For the *Pinterest* and *Beauty* datasets, the recommendation performance is insensitive to the number of heads, possibly because user preference has little effect.

Figure 13 shows the recommendation performance for the proposed approach using the three datasets, where the length of time interval (i.e., window size) ranges from 2 to 7 (**RQ4**). As shown in figure 13, for the *Pinterest* datasets, the recommendation performance significantly increases as the length of the time interval increases if the window size is small, and is insensitive to the length of the time interval if the window size is large. This shows that a pin that is clicked by the user is more related to pins with which there has been a recent interaction by the same user in a given time interval. Similarly, for the *Beauty* datasets, the recommendation performance significantly increases as the length of the time interval increases. This shows that a beauty product that is clicked by the user is more related to beauty products with which the user has recently interacted within a given time interval. For the *MovieLens-100k* dataset, the recommendation performance slightly decreases as the length of time interval increases if the window size is small, and increases as the length of time interval increases when the window size is larger than 6. This shows that the movie that is chosen by a user is less related to the movies with which the user has recently interacted within a given time interval, but is more related the user preferences for type of movie. The recommendation performance increases as the length of the time interval increases, because changing preferences are easy identified over a long time interval. This demonstrates that the proposed approach identifies changes in the user's preferences. When users interact with items that are different from those they have seen before, the proposed module adapts and recommends an item more accurately.

TABLE IV
TOP-K RECOMMENDATION PERFORMANCE FOR K = 10 FOR THE PROPOSED MODEL, WITH AND WITHOUT POSITIONAL ENCODING (PE)

| | | HR@10 | MRR@10 | NDCG@10 |
|---|---|---|---|---|
| MovieLens-100k | Remove PE | 0.7338 | 0.3681 | 0.4545 |
| | With PE | **0.7603** | **0.3861** | **0.4747** |
| Pinterest | Remove PE | **0.9064** | **0.5668** | **0.6491** |
| | With PE | 0.9046 | 0.5619 | 0.645 |
| Beauty | Remove PE | 0.5542 | 0.2348 | 0.3097 |
| | With PE | **0.5587** | **0.2382** | **0.3132** |

Table IV shows the top-*k* recommendation performance for *k* = 10 for the proposed model, with and without positional encoding (PE). As shown in Table IV, the positional encoding strategy improves the recommendation performance for the *MovieLens-100k* and *Beauty* datasets but the recommendation performance for the *Pinterest* dataset is better when the positional encoding strategy is not used in the proposed algorithm. Without the strategy of positional encoding in the proposed contextual item attention modules, the attention weight for each item depends only on the item embedding. The model makes recommendations based on the items with which a user has recently interacted, but does not consider the order of interaction for items. Positional encoding is not required for sparse datasets, where user-item interaction lists are usually short (e.q., *Pinterest* dataset). Recommendation system with positional encoding performs better on denser datasets because the order of items with which the user has recently interacted items within a given time interval is considered.

The experimental results show that the characteristics of each dataset are different. In terms of the *MovieLens-100k* dataset, movie types and user tastes are complex, so a deeper stacked structure is required for the multi-head user attention module to identify fundamental user preferences. A wider multi-head user attention module with more heads is also required to identify more aspects of user preferences for movies. However, the width and depth of multi-head user attention module does not significantly affect the recommendation performance for *Pinterest* and *Beauty* datasets.

The recommendation performance for the *Pinterest* and *Beauty* datasets is affected by the length of the time interval. The recommendation performance increases significantly as the length of the time interval increases, so the target item is more related to items with which the user has recently interacted within a given time interval for *Pinterest* and *Beauty* datasets. The recommendation performance for *MovieLens-100k* and *Beauty* datasets is better if a positional encoding strategy is used for the proposed algorithm. The recommendation performance for the *Pinterest* dataset is better if positional encoding strategy is not used by the proposed algorithm. Real recommendation applications are complex so the proposed recommendation algorithm has a high degree of flexibility and adapts to the characteristics of recommendation applications by setting different depths, widths and lengths of time interval for the model. The proposed algorithm achieves performs better in real recommendation applications than current state-of-the-art methods.

## V. CONCLUSIONS

Users' changing preferences can significantly affect the accuracy of a recommendation system so users' preferences must be modeled and the patterns of users' changing preferences is necessary must be captured. This can be difficult for a recommendation system. This study proposes a novel context-aware recommendation system that adapt to users' changing preferences. Two different attention mechanisms are used. An attention mechanism is used for temporal contextual information about an item for the proposed contextual item attention module and user-item attention is used to learn the attention weight between several items within a defined time interval. A multi-head attention mechanism is used for the proposed multi-head user attention module to capture different aspects of user preferences. A stacked structure is used for the multi-head user attention module to capture different aspects of user preferences repeatedly and to identify fundamental preferences. Finally, fixed positional encoding gives additional temporal information.

The experiment compares the proposed model with several state-of-the-art recommendation baseline methods. The results show that the proposed model with a stacked structure significantly outperforms the baseline methods. The additional temporal information also allows the model to perform well with benchmark datasets.

Deep learning-based methods better mode users' and items' representation and capture users' preferences. A deep learning based attention mechanism allows the proposed model to capture users' dynamic preferences more precisely. Using contextual information and additional temporal information, the recommendation system assesses users better and provides more suitable recommendations for items.

Future work will add other contextual information for the recommendation, such as topics or categories of items within defined time intervals, to enhance the ability to adapt to users' changing preferences. Users' changing preferences will also be formulated as a multi-armed bandit problem, which is solved using multi-armed bandit methods and Reinforcement Learning (RL) methods. Using multi-armed bandit methods and Reinforcement Learning (RL) methods, any change in preferences can be quantified and the learned strategy adapted.

## Acknowledgment

This research work was supported in part by the Ministry of Science and Technology Research Grant MOST 109-2634-F-006-016 -.